\documentclass[prl,aps,twocolumn,showpacs]{revtex4}

\usepackage{graphicx}
\usepackage{amsmath,amsfonts,amsbsy,amssymb}
\usepackage{mathrsfs,bbm}

\newcommand{\dual}[1]{\langle #1\rvert}
\newcommand{\ket}[1]{\lvert#1\rangle}

\newcommand{\op}[2]{\ket{#1}\dual{#2}}
\newcommand{\proj}[1]{\ket{#1}\dual{#1}}

\begin{document}
\title{Collective States and Symmetric Local Decoherence in Large Ensembles of Qubits}
\author{Bradley A. Chase}
\email{bchase@unm.edu}
\author{JM Geremia}
\email{jgeremia@unm.edu}
\affiliation{Department of Physics \& Astronomy, The University of New Mexico, Albuquerque, New Mexico 87131 USA}
\begin{abstract}
The \textit{symmetric collective states} of an atomic spin ensemble (i.e., many-body states that are invariant under particle exchange) are not preserved by decoherence that acts identically but individually on members of the ensemble.  We develop a class of collective states in an ensemble of $N$ spin-1/2 particles that is invariant under symmetric local decoherence and find that the dimension of the Hilbert space spanned by these collective states scales only as $N^2$.  We then investigate the open system dynamics of experimentally relevant non-classical collective atomic states, including Schr\"{o}dinger cat and spin squeezed states, subject to various symmetric but local decoherence models.
\end{abstract}
\date{\today}

\pacs{03.65.Fd,03.65.Yz,34.10.-x}

\maketitle

Many fields, including precision metrology \cite{Wineland1993,Romalis2003}, quantum information science \cite{Kuzmich2003,Polzik2001,Black2005,Chaudhury2007} and quantum optical models of condensed matter phenomena \cite{Greiner2002,Sadler2006,Morrison2008}, utilize collective spin degrees of freedom in an ensemble of atoms or ions.  An attractive feature of such ensembles is that one can achieve exquisite control over their quantum dynamics, even in practice \cite{Polzik2001,Black2005,Chaudhury2007}.  A second benefit is that interesting many-body effects can be realized within a small, theoretically manageable, sub-Hilbert space of the entire spin ensemble's state space.   Even though the joint Hilbert space $\mathscr{H}_N$ for $N$ spin-$j$ particles is immense, $\dim(\mathscr{H}_N) = (2j+1)^N$, it is not uncommon to find dynamical symmetries that reduce the effective dimension of the ensemble without sacrificing the rich structure characteristic of many-body interactions and entanglement.

Most previous work on dynamical symmetries in spin ensembles has involved the \textit{symmetric collective states} \cite{Stockton2003}: the sub-Hilbert space $\mathscr{H}_\mathrm{S} \subset \mathscr{H}_N$ spanned by $N$-body states that are invariant under the permutation of any two particle labels $\hat{\Pi}_{ij} \ket{\psi} = \ket{\psi}$, $\ket{\psi} \in \mathscr{H}_\mathrm{S}$.   The dimension of this so-called ``symmetric space'' grows only linearly in the number of spin particles  $\dim(\mathscr{H}_\mathrm{S}) = (2 N j+1) \ll (2 j+1)^N $, yet exhibits phenomena of considerable interest, such as spin-squeezing \cite{Kitagawa1993,Hald1999} and zero-temperature phase transitions \cite{Morrison2008}.   For $\mathscr{H}_\mathrm{S}$ to provide an accurate description of the ensemble's state, the dynamics must be generated by completely symmetric collective processes--- \emph{symmetric} indicating that the processes are permutation invariant, and \emph{collective} meaning that they can be expressed in terms of total angular momentum operators.  Symmetric states are accessed in practice by experiments that avoid particle addressability, such as by homogeneous coupling to the external fields used to control and observe the ensemble \cite{Stockton2003,Chaudhury2007}.  

Unfortunately many of the decoherence models most appropriate for large spin ensembles cannot be described as collective processes even when the decoherence acts identically on each particle.  The $(N+1)$-dimensional symmetric Hilbert space $\mathscr{H}_\mathrm{S}$ is not preserved under such decoherence, greatly complicating dynamical simulations of the ensemble.  It has thus become common practice either to: (1) restrict the dynamics to very short-times when decoherence can be approximately neglected; or (2) model the decoherence as a collective process, even when doing so is not necessarily  physically justified.  In atomic spin ensembles, for example, a typical source of decoherence comes from spontaneous emission, yet collective radiative processes only occur under specific conditions such as superradiance from highly confined atoms \cite{Dicke1954} and some cavity-QED or spin-grating settings \cite{Black2005}.  In many ongoing experiments,  fluorescence imaging can, to varying degree,  distinguish between particles.

In this Letter, we generalize the collective states of an ensemble of spin-1/2 particles (qubits) to include states that are preserved under symmetric--- but not necessarily collective--- transformations.      Specifically, we generalize from the strict condition of complete permutation invariance to the broader class of states that are indistinguishable across degenerate irreducible representations (irreps) of the rotation group. The representation theory of the rotation group has, of course, been utilized in a wide variety of contexts, such as to protect quantum information from decoherence by encoding it into degenerate irreps with the same total angular momentum \cite{Bacon2001, Lidar1998}.   Here, we utilize relevant aspects of the representation theory to obtain a reduced-dimensional description of quantum maps that act locally but identically on every member of an ensemble of qubits.  The dimension of the Hilbert space $\mathscr{H}_\mathrm{C}$ spanned by our generalized collective states scales favorably, $\dim(\mathscr{H}_\mathrm{C}) \sim N^2$, making it possible to address decoherence models relevant to a greater variety of experimental settings.  To demonstrate the viability of our approach, we investigate several decoherence models applied to Schr\"{o}dinger cat and spin-squeezed states of an atomic ensemble.

\textit{Representations of the Rotation Group.---} We consider an ensemble of $N$ identical spin-1/2 particles, each described by the single-particle angular momentum operators $\hat{\sigma}^{(n)} = ( \hat{\sigma}_\mathrm{x}^{(n)}, \hat{\sigma}_\mathrm{y}^{(n)}, \hat{\sigma}_\mathrm{z}^{(n)})$.  The joint Hilbert space for the entire spin ensemble $\mathscr{H} = \mathscr{H}^{(1)} \otimes \cdots \otimes \mathscr{H}^{(N)}$ has dimension $\dim( \mathscr{H}) = 2^N$, and arbitrary pure states of the ensemble can be expressed in the tensor product basis $\ket{\psi} = \sum_{m_n} c_{m_1,\ldots,m_N} \ket{m_1,m_2,\ldots,m_N}$ where the $\ket{m_1,\ldots,m_N} = \ket{\frac{1}{2},m_1}_1 \otimes \cdots \otimes \ket{\frac{1}{2},m_N}_N$ satisfy $\hat{j}^{(n)}_\mathrm{z} \ket{m_1,\ldots,m_N} =  m_n \ket{m_1,\ldots,m_N}$.

Each particle in the ensemble transforms separately under a rotation such that $\ket{\psi'} =  [\mathscr{D}^{\frac{1}{2}}(R)]^{\otimes N} \ket{\psi}$,  where $\mathscr{D}^\frac{1}{2}(R)$ is the spin-1/2 rotation operator parameterized by the Euler angles $R=(\alpha,\beta,\gamma)$.  Expressed in the tensor-product basis, the $[\mathscr{D}^{\frac{1}{2}}(R)]^{\otimes N}$ provide a reducible representation for the rotation group but can be decomposed into irreducible components (irreps)
\begin{equation} \label{Equation::IrrepDecomp}
	\mathscr{D}(R) = \bigoplus_{J=J_\mathrm{min}}^{J_\mathrm{max}} 
	 \bigoplus_{i=1}^{d^J_n} \mathscr{D}^{J,i}(R) 
\end{equation}
via the total spin eigenstates $\hat{J}^2 \ket{J,M,i}  =  J(J+1) \ket{J,M,i}$ and $\hat{J}_\mathrm{z} \ket{J,M,i}  =  M \ket{J,M,i}$, with $\hat{J}_q = \frac{1}{2} \sum_{n=1}^N \hat{\sigma}_q^{(n)}$ and $J=\mathrm{mod}(\frac{N}{2},2),\ldots, \frac{N}{2}$.  For each total angular momentum $J$, the quantum number $i=1,\ldots,d^J_N$ distinguishes between the 
\begin{equation}
   	 d^J_N =\frac{N!(2J+1)}{(N/2-J)!(N/2+J+1)!} 
\end{equation}
degenerate irreps with total angular momentum $J$ \cite{Mikhailov1977}.  

\textit{Collective States.}---  In the ``irrep basis,'' arbitrary pure states of the spin ensemble are expressed as $\ket{\psi(t)} = \sum_{J,M,i}  c_{J,M,i}(t) \ket{J,M,i}$,  which still requires $2^N$ coefficients.  Of course, simply transforming to the irrep basis does not change the effective dimension of the Hilbert space, but it suggests a symmetry that can be exploited to do so.  We generalize the definition of \textit{collective states} to the sub-Hilbert space $\mathscr{H}_\mathrm{C}$ spanned by $N$-particle states $\ket{\psi_\mathrm{C}}$ that are indistinguishable across the $d_N^J$ degenerate irreps for each total angular momentum $J$.  This introduces the symmetry $c_{J,M,i} = c_{J,M,i'}, \forall i,i'$, making it unnecessary to distinguish $\ket{J,M,i}$ over irrep label.  Defining effective kets $\ket{J,M}$ on each total-$J$ irrep, we then have $\ket{\psi_\mathrm{C}} = \sum_{J,M} c_{J,M} \ket{J,M}$ with the coefficients
\begin{equation}
		c_{J,M}(t) = \sqrt{\frac{1}{d^J_N}} \sum_{i=1}^{d^J_N} c_{J,M,i}(t) \,.
\end{equation}
Under this symmetry, $\dim(\mathscr{H}_\mathrm{c}) = (N+2)^2/4$ (for $N$ even) scales only quadratically with the number of particles  and therefore admits efficient representation.  When describing decoherence and other open-system dynamics, such as continuous measurement, one may require the \textit{collective state density operator}, which we define as $\hat{\rho}_\mathrm{C} = \oplus_J \hat{\rho}_J$ to give
\begin{equation} \label{Equation::RhoReduced}
	\hat{\rho}_\mathrm{C}(t)= \sum_J \sum_{M,M'} \rho_{J,M;J,M'}(t) \op{J,M}{J,M'} .
\end{equation}
These collective density operators restrict against coherences between irreps \cite{note1}, but still subsume the symmetric space previously considered for large spin ensembles \cite{Stockton2003}.   $\mathscr{H}_\mathrm{S} \subset \mathscr{H}_\mathrm{C}$ is spanned by the maximal angular momentum manifold: $c_{J,M}=0$ for $J \neq N/2$.   All $\ket{\psi_s} \in \mathscr{H}_\mathrm{S}$ can be obtained from the permutation-invariant state $\ket{N/2,N/2} \leftrightarrow \ket{\frac{1}{2},\ldots,\frac{1}{2}}$ by a simple rotation $\mathscr{D}^{N/2}(R)$ and are collective states since $d^{N/2}_N = 1$.

\begin{figure*}[th!]
\begin{center} \includegraphics[scale=1]{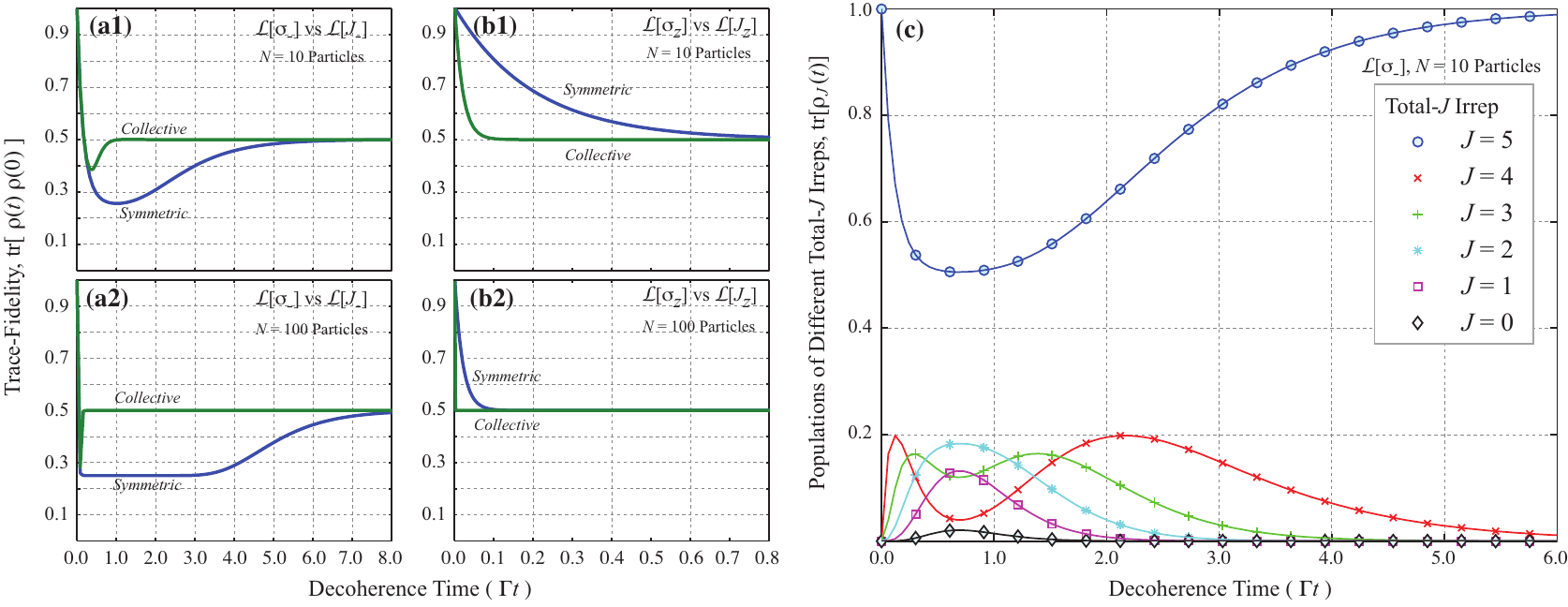} \end{center}
\vspace{-5mm}
\caption{(color online) Decoherence of an initial Schr\"{o}dinger cat state: (a1-a2) time-dependent fidelity with the initial state for both symmetric local $\mathcal{L}[\hat{\sigma}_-]$ and collective $\mathcal{L}[\hat{J}_-]$ decoherence for different numbers of particles; (b1-b2) similar comparison for $\mathcal{L}[\hat{\sigma}_\mathrm{z}]$ versus $\mathcal{L}[\hat{J}_\mathrm{z}]$; (c) time-dependent populations of different total-$J$ irreps for $\mathcal{L}[\hat{\sigma}_-]$. \label{Figure::catStateDecoherence}}
\end{figure*}

\textit{Collective State-Preserving Processes.}--- We next discuss quantum maps that preserve collective states $\mathcal{L} : \mathscr{H}_\mathrm{C} \rightarrow \mathscr{H}_\mathrm{C}$.  The super-operator
\begin{equation} \label{Equation::LCollective}
	\mathcal{L} \hat{\rho}_\mathrm{C} = 
		\sum_{J} \sum_{M,M'} \rho_{J,M;J,M'} f^J_{MM'}
\end{equation}
will be closed on $\mathscr{H}_\mathrm{C}$ if its action $f^{J}_{MM'} = \mathcal{L} \op{J,M}{J,M'}$ can be expressed in the irrep basis in such a way that it does not distinguish between degenerate irreps.  Operators that transform simply with respect to the rotation group $\hat{S} = \sum_{J} \sum_{M,M'} s^{J}_{M,M'}\op{J,M}{J,M'}$, including the \textit{collective angular momentum operators} $\hat{J}_i$ and all collective operators $\hat{S} = \sum_{n=1}^N \hat{s}^{(n)}$ formed from $\hat{s}^{(n)}\in \mathfrak{su}(2)$, satisfy this requirement directly.  

Unfortunately, not all dynamics can be expressed entirely in terms of collective operators; they might only be symmetric over local single-particle super-operators, 
\begin{equation} \label{Equation::SymmetricL}
	\mathcal{L}[ \hat{s} ] \hat{\rho} = \sum_{n=1}^N \mathcal{L}^{(n)} [ \hat{s}^{(n)} ] \hat{\rho},
\end{equation}
where $\mathcal{L}^{(n)}[ \hat{s}^{(n)} ]\hat{\rho} = \hat{s}^{(n)} \hat{\rho} (\hat{s}^{(n)})^\dagger$ for $\hat{s}^{(n)} \in \mathfrak{su}(2)$.  Our main result is that any symmetric local map of the form Eq.\ (\ref{Equation::SymmetricL}) can be brought into the form of Eq.\ (\ref{Equation::LCollective}) and therefore preserves collective states.  For the $\mathfrak{su}(2)$ operators $\hat{s} = \vec{s} \cdot \hat{\sigma}$ and $\hat{s}^\dagger= \vec{t} \cdot \hat{\sigma}$, $f^{J}_{MM'}$ can be constructed as $f^{J}_{MM'} = \vec{s} \cdot \mathbf{g}(J,M,M') \cdot \vec{t}$ from the tensor $g_{qr} =  \sum_{n=1}^N \hat{\sigma}_{q}^{(n)}\op{J,M}{J,M'}(\hat{\sigma}_{r}^{(n)})^{\dagger}$ whose elements we find to take the form \cite{Chase2008b}
\begin{widetext}
\begin{eqnarray}
g_{qr}(J,M,M') & = & \frac{1}{2} \left[   \frac{A_q^{J,M} A_r^{J,M'}}{J} \left( 1+\frac{\alpha^{J+1}_N(2J+1)}{d^J_N(J+1)}\right)
		\op{J,M_q}{J,M'_r}  + \frac{B_q^{J,M}B_r^{J,M'}  \alpha^J_N}{d^J_N J}\op{J-1,M_q}{J-1,M'_r}  
	\right. \nonumber
		 \\ 
	&& \left. \quad \quad \quad + \frac{\alpha^{J+1}_N D_q^{J,M} D_r^{J,M'} }{d^{J}_N (J+1)}
		  \op{J+1,M_q}{J+1,M'_r}  \right], 
		  \label{Equation::MainResult}
\end{eqnarray}
\end{widetext} 
where the indices $q,r \in \{-,+,z,0\}$ denote representing single-particle angular momentum operators in the basis $\{ \hat{\sigma}_-, \hat{\sigma}_+, \hat{\sigma}_\mathrm{z}, \hat{1}\}$.  We have also defined the following symbols: $M_+ = M + 1$, $M_- = M - 1$ and $M_z = M$, the reduced degeneracies $\alpha^J_N = \sum_{J'=J}^{\frac{N}{2}}d^{J'}_N$, and the tensor coefficients
\begin{eqnarray}
			A_\pm^{J,M} &= & + \sqrt{(J \mp M)(J \pm M+1)}\\
			A_z^{J,M} &= & M \\
			B_\pm^{J,M} &= & \pm \sqrt{(J \mp M)(J \mp M-1)}\\
			B_z^{J,M} &= & \sqrt{(J+M)(J-M)} \\
			D_\pm^{J,M} &= & \mp \sqrt{(J \pm M+1)(J \pm M+2)} \\
			D_z^{J,M} &= &\sqrt{(J+M+1)(J-M+1)} \ .
\end{eqnarray}
The three terms in Eq.\ (\ref{Equation::MainResult}) arise from two types of processes: (Term 1) transitions that occur between $M$ levels within a single $J$ irrep; and (Terms 2-3) transitions that couple neighboring irreps with $\Delta J = \pm 1$.   For brevity, we have not listed terms in Eq.\ (\ref{Equation::MainResult}) for $q=0$ (the identity operator); these terms as well as those listed here can be determined following the methods in Ref.\ \cite{Chase2008b}.
  
Reducing the effective dimension of the ensemble Hilbert space to $O(N^2)$ should be adequate to allow simulations of $N > 100$ on a desktop computer, and much more on specialized hardware.  It is, however, worth mentioning that the structure of the mapping in Eq.\ (\ref{Equation::MainResult}) suggests an opportunity for further reducing the effective dimension of the ensemble's Hilbert space.  Since the dynamics only couple neighboring irreps, if the initial state is confined to only a single irrep (for example a symmetric state)  then it may well be possible to truncate distant irreps from the decomposition without introducing appreciable error into the ensemble dynamics.

\textit{Modeling Decoherence.}--- As discussed in the introduction, realistic decoherence models for an ensemble of spin particles are often described most aptly by a symmetric sum over local channels.  Consider, for example, the open system dynamics governed by the master equation
\begin{equation} \label{Equation::master}
	\frac{d \hat{\rho}(t)}{dt}  = - i [ \hat{H}, \hat{\rho}(t) ] + \Gamma \mathcal{L}[ \hat{s} ]
		\hat{\rho}(t),
\end{equation}
where $\hat{H}$ (and any measurements performed) are described by collective operators, but the decoherence involves the symmetric Linblad superoperator
\begin{equation} \label{Equation::SymmetricDecoherence}
	\mathcal{L}[ \hat{s} ] \hat{\rho} =
		\sum_{n=1}^N  \hat{s}^{(n)} \hat{\rho} [\hat{s}^{(n)}]^\dagger  
		- \frac{1}{2}
		\left( [\hat{s}^\dagger \hat{s}]^{(n)} \hat{\rho} +
		\hat{\rho} [\hat{s}^\dagger \hat{s}]^{(n)}  \right)
\end{equation}
of the form in Eq.\ (\ref{Equation::SymmetricL}).  As this decoherence model does not preserve symmetric states, it has been common practice to consider instead the associated collective process
\begin{equation} \label{Equation::CollectiveL}
	\mathcal{L}[ \hat{S} ] \hat{\rho} \equiv  \left[
		\hat{S} \hat{\rho} \hat{S}^\dagger - \frac{1}{2} \left(
			\hat{S}^\dagger \hat{S} \hat{\rho} + \hat{\rho}  \hat{S}^\dagger \hat{S} \right) \right]
\end{equation}
with $\hat{S} = \sum_n \hat{s}^{(n)}$.

To illustrate the difference between symmetric and collective decoherence models, we considered the open system dynamics of two representative problems.  First, we compared the dynamics generated by the symmetric-local $\mathcal{L}[\hat{s}]$ versus collective $\mathcal{L}[\hat{S}]$ Linblad master equations applied to an initial Schr\"{o}dinger cat state $\ket{\psi(0)} = \left( \ket{\frac{N}{2},+ \frac{N}{2}} 
	                    + \ket{\frac{N}{2},-\frac{N}{2}} \right) / \sqrt{2}$. 
Figure \ref{Figure::catStateDecoherence}(a-b) depicts the fidelity $\mathcal{F}(t) =  \dual{\psi(0)} \hat{\rho}(t)\ket{\psi(0}$ evolved under Eq.\ (\ref{Equation::master}) (with $\hat{H} = 0$) for two different types of decoherence channels: Fig.\ \ref{Figure::catStateDecoherence}(a1-a2) compares the collective versus symmetric master equations with $\hat{s} = \hat{\sigma_z}$ for $N=10$ and $N=100$ particles, respectively; and Fig.\ \ref{Figure::catStateDecoherence}(b1-b2) makes a similar comparison for $\hat{s} = \hat{\sigma}_\mathrm{z}$.  The examples we considered (including some not reported here) suggest symmetric local decoherence models can generate dynamics that are appreciably different from their collective analogs.  This is perhaps not too surprising: for an initially symmetric state, collective decoherence models $\mathcal{L}[ \hat{S}]$ confine the dynamics to only maximum-$J$ irrep; symmetric local models $\mathcal{L}[\hat{s}]$ do not necessary preserve the irrep decomposition of the initial state.  Fig. \ref{Figure::catStateDecoherence}(c) depicts the norm of each total-$J$ irrep block of the density operator $N_J = \mathrm{tr}[ \hat{P}_J \hat{\rho}(t)]$ as a function of time for $\mathcal{L}[ \hat{\sigma}_-]$ ($\hat{P}_\mathrm{J} = \sum_M \proj{J,M}$).   The observation that small-$J$ irreps are only minimally populated suggests that further model reduction by truncating the Hilbert space to only the largest $J$ blocks could be beneficial.

\begin{figure}
\begin{center}	\includegraphics{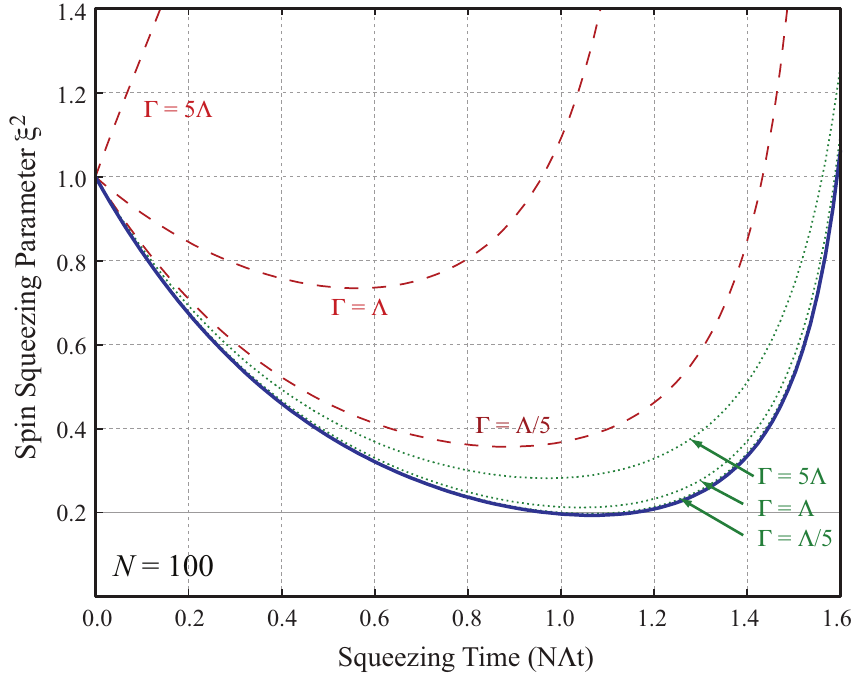} \end{center}
\vspace{-5mm}
\caption{(color online) Time-evolution of the squeezing parameter $\xi^2$ for a spin ensemble driven by $\hat{H} = -i  \Lambda ( \hat{J}_+^2 - \hat{J}_-^2)$ subject to $\mathcal{L}[ \hat{\sigma}_-]$ (dotted lines) and $\mathcal{L}[ \hat{J}_-]$ (dashed lines) with relative decoherence rates $\Gamma = \Lambda/ 5, \Lambda, 5 \Lambda$.  For comparison, the solid line denotes decoherence-free squeezing.
	\label{Figure::Squeezing}
}
\end{figure}

As a second example, we compared symmetric-local versus collective decoherence models applied to dynamically-generated spin squeezing under the counter-twisting Hamiltonian $\hat{H} = -i \Lambda (\hat{J}_{+}^2 - \hat{J}_{-}^2)$ \cite{Kitagawa1993}.  We performed simulations by time-evolving Eq.\ (\ref{Equation::master}) from the initial spin-coherent state $\ket{\frac{N}{2}, \frac{N}{2}}$ for $N=100$ with $\mathcal{L}[\hat{\sigma}_-]$ and $\mathcal{L}[\hat{J}_-]$.  Figure \ref{Figure::Squeezing} depicts the time-dependent squeezing parameter $\xi^2 = N \langle \Delta \hat{J}_y^2 \rangle / \langle \hat{J}_z \rangle ^2$, each for $\Gamma = \Lambda/5, \Lambda, 5 \Lambda$.   Under the conditions we considered, symmetric local decoherence wave evidently less destructive to the squeezing dynamics than collective models.  As observed for the cat-state dynamics, to large extent the main effect of symmetric-local decoherence is leakage from the maximum $J$ irrep.  But since the driving Hamiltonian $\hat{H}$ involves only collective spin operators, the coherent dynamics decouple for different total $J$: the population in each irrep block then undergoes its own squeezing, evidently making the dynamics more resistant to symmetric local decoherence than collective processes.

\textit{Conclusion.---} We have presented an exact reduced-dimensional description of quantum maps that act locally but identically on each member of an ensemble of qubits.  Our results confirm that the common practice of modeling decoherence in large atomic ensembles via a collective process (when not physically motivated) is unlikely to allow for quantitative comparison between theory and experiment.  The techniques developed here are expected to help achieve such agreement, and thus play an important role in quantum measurement, control and information theoretic settings where high-fidelity modeling is essential to interpreting experimental data.    This work was inspired by discussions with Ivan Deutsch and Poul Jessen and was supported by the NSF (PHY-0652877) and the DOE under contract with Sandia National Laboratory through the NINE program (707291).


\end{document}